 \documentclass[final,5p,times,twocolumn]{elsarticle}

\usepackage{graphicx}
\usepackage{amssymb}

\usepackage{lineno}
\usepackage{listings}
\usepackage{adjustbox}
\usepackage{caption} 
\usepackage{hyperref}
\modulolinenumbers[5]

\usepackage{xcolor}

\usepackage{amssymb}

\usepackage{longtable}
\usepackage{booktabs}
\usepackage{array}
\newcolumntype{C}[1]{>{\centering\arraybackslash}m{#1}}
\newcolumntype{R}[1]{>{\raggedleft\arraybackslash}m{#1}}

\journal{Journal of Computer Standards \& Interfaces}

\begin{document}

\begin{frontmatter}

\title{A Community-Driven Validation Service for Standard Medical Imaging Objects}



\author{Jorge Miguel Silva,Tiago Marques Godinho,David Silva,Carlos Costa }%
\address{DETI/IEETA, University of Aveiro, Portugal}

\begin{abstract}
Digital medical imaging laboratories contain many distinct types of equipment provided by different manufacturers. Interoperability is a critical issue and the DICOM protocol is a de facto standard in those environments. However, manufacturers' implementation of the standard may have non-conformities at several levels, which will hinder systems' integration. Moreover, medical staff may be responsible for data inconsistencies when entering data. Those situations severely affect the quality of healthcare services since they can disrupt system operations. The existence of software able to confirm data quality and compliance with the DICOM standard is important for programmers, IT staff and healthcare technicians. Although there are a few solutions that try to accomplish this goal, they are unable to deal with certain situations that require user input. Furthermore, these cases usually require the setup of a working environment, which makes the sharing of validation information more difficult.
This article proposes and describes the development of a Web DICOM validation service for the community. This solution requires no configuration by the user, promotes validation results’ share-ability in the community and preserves patient data privacy since files are de-identified on the client side.

\end{abstract}

\begin{keyword}
Web \sep DICOM \sep PACS \sep Medical Imaging \sep Validator

\end{keyword}

\end{frontmatter}

\section{Introduction}
\label{S:1}
In recent decades, healthcare institutions have been continuously increasing the production of digital medical imaging data. In part, this was  due the increase of digital medical imaging equipment and information systems, which are now fundamental in medical diagnosis, decision support and treatment procedures. Picture Archiving and Communication System (PACS) is predominant in this field, providing tools for data acquisition, storage, distribution and visualization. It is a mature concept supported by a set of hardware and software technologies, being grounded in the Digital Imaging and Communications in Medicine (DICOM) standard to ensure normalized data formats and processes. It is a universally accepted standard in medical imaging laboratories, designed to encompass all functional aspects~\cite{Huang_2010,Creighton1999ASystems,St.Cyr2013AnStandards}. Nowadays, the communication between equipment and information systems is usually done using the DICOM standard~\cite{Mildenberger2002IntroductionStandard}. This defines the reference information model, how data is encoded and communicated. Data is merged in structured objects that follow normalized templates per image modality, which contain metadata related with the procedure, patient, acquisition technique and institution, besides pixel data.

Regular workflows are so supported by PACS  ~\cite{Webb2003IntroductionImaging,2003RealizationSystem}, that the existence of non-conforming applications or equipment may disrupt the regular operation with potential losses in the medical undertaking~\cite{dreyer2006pacs}. 

Despite the existence of DICOM standard, the reality is that challenges to interoperability still arise. Furthermore, technology is constantly evolving and DICOM needs to be updated, thus, hindering compliance between equipment.

The baseline to ensure interoperability between different systems is the DICOM Conformance Statement, since it provides a foundation to determine connectivity and assess the potential interoperability of two products. In some cases, it is possible to identify potential problems without ever having the products physically connected. It is a public document that must be provided by the vendor which describes the DICOM capabilities and functions implemented in a product, allowing connectivity comparisons and defining all the necessary information to perform a certain functionality \cite{dicomPart2Conformance}. 
DICOM validation software is important to assist in the testing of products’ DICOM conformance, providing an independent measurement of the accuracy of products’ DICOM interface.

Notwithstanding, verifying the compliance of data produced by PACS applications is not trivial, since the DICOM standard supports a significant variety of modalities and information entities, each one with its specifications and dependencies. The intrinsically complex nature of this scenario motivates the development of tools and methodologies capable of testing the compliance of produced DICOM objects with the standard.
This article proposes and describes the development of a Web DICOM validation service for the medical imaging community that agglutinates, in a unique way, a set of functionalities. Its use can be as simple as uploading DICOM objects to be checked, without requiring platform registration or authentication, but ensuring data privacy by removing the patient’s personal health information on the client side. Then, more complex validation tasks can be performed by the community in a collaborative way.

\vfill

\section{DICOM Constitution}
\label{S:2}
\begin{figure}[!ht]
  \centering
    \includegraphics[]{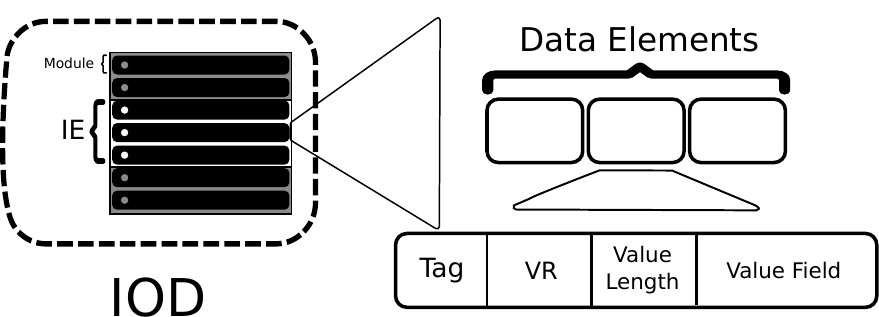}
 	\caption{\label{fig:IOD_struct} Constitution of an IOD}
\end{figure}

DICOM Information Model (DIM) rules the organization of information structures in the standard. It specifies the relationship between DICOM objects' information entities (IE) and real-world entities such as the patient, study, series and image. IE are computer data model abstractions for the real-word objects. Each IE may contain one or more Modules that are basic aggregations of related Data Elements (or Attributes). For instance, the patient module contains the name, ID, birth-date and sex attributes~\cite{Horii2011DICOM}. 
An object template is denominated as DICOM Information Object Definitions (IODs) and may contain one or more IE. IODs are normalized collections of DICOM Data Elements organized in Modules and IE (Figure~\ref{fig:IOD_struct}). The IOD Modules may be mandatory, conditional or user optional, as described in the DICOM Standard Part 3 \cite{Rosslyn2016}. 
Data Elements follow a TLV (Tag-Length-Value) encoding schema according to Part 5 of the standard \cite{Nema2018DigitalEncoding}. The DICOM tag identifies an attribute using two hexadecimal numbers, called the group and element respectively.  These numbers are specified as the ordered pair (\(<group>,<element>\)). For instance, Tag (0010,0010) identifies the Patient Name element of the Patient group. Length defines the size of the attribute’s Value. The Value field contains the attribute’s data. According to DICOM transfer syntax, an optional Value Representation (VR) element may also be present and specifies the attribute data type. There is also a Value Multiplicity that specifies the number of values that can be encoded in the value field of that Data Element~\cite{Oosterwijk2005DICOMBasics}. 
The list of normalized Data Elements is defined in the DICOM Dictionary available in part 6 of the standard ~\cite{dicom6} and, according to the presence in Modules, are classified as:

\begin{itemize}
\item \textbf{Type 1}: Attribute presence is mandatory and must have a valid value;
\item \textbf{Type 2}: Attribute presence is mandatory, but its value may be left blank;
\item \textbf{Type 3}: Attribute presence is optional.
\end{itemize}

Furthermore, all types of attributes can be conditional (C), since IODs and Service-Object Pair (SOP) Classes, a combination of a DICOM service command (DIMSE) and an IOD, can define Data Elements that shall be included under certain specified conditions.
Conditional types have the same requirements as their type (1, 2 or 3) under these conditions. As such, it is a protocol violation if the specified conditions are met, and the Data Element is not included. On the other hand, when the specified conditions are not met, Type C elements shall not be included in the dataset \cite{Nema2018DigitalEncoding}.

\section{Related Work}
\label{S:3}
DICOM IODs are flexible structures and verification of objects' compliance with standard definitions may be a complex task. Due to the need to ensure the robustness and accuracy of software applications, programmers were the first to feel the need for verification tools. Then, healthcare IT staff requested end-user software applications to confirm the conformity of enterprise DICOM network nodes and debug abnormal events.

\textit{DCMCHECK}\footnote{DCMCHECK: http://dicom.offis.de/dcmcheck.php.en} is commercial software that tries to solve this issue. It uses a specialized IOD description language which allows extensions (e.g. private elements, DICOM correction proposals) to be added to the IOD definition without changing the application itself ~\cite{Hewett1997ConformanceObjects}. The DICOM files are verified as conforming to the standard IOD definition (DICOM Part 3~\cite{Rosslyn2016})]), data structures and encoding (DICOM Part 5~\cite{Nema2018DigitalEncoding}) and the data dictionary (DICOM Part 6 ~\cite{dicom6}). Furthermore, the DICOM File Meta Information (Preamble + DICOM Prefix + File Meta Information (0002, xxxx)) is evaluated according to the DICOM Part 10 specifications ~\cite{NEMA2018DigitalInterchange}, as well as the consistency between it and the rest of the DICOM meta data information on the file.

DICOM Validation Toolkit (DVTk)\footnote{DVTk: https://www.dvtk.org/} is an open source project for testing, validating and diagnosing problems with communication protocols in medical imaging environments \cite{Potter2007}. It supports DICOM, HL7 and IHE integration profiles, and provides a DICOM Attribute Validator for validating DICOM files against definition files. The validator application includes GUI and command line versions, and a collection of .NET libraries for creating new validation and test tools. Moreover, it provides a DICOM Attribute Validator for validating DICOM files against definition files.

There are also other examples of open-source validation software like, for instance, the \textit{dicom3tools/dciodvfy}\footnote{dicom3tools/dciodvfy: http://www.dclunie.com/dicom3tools/dciodvfy.html} and the \textit{dcm4che3} validator \footnote {dcm4che3: https://github.com/dcm4che/dcm4che}. In general, they can check for inconsistencies in the DICOM files against Part 3 and 5 of the standard, Multiplicity against the Data Dictionary and Data Element Value content against encoding rules defined by the standard. 
Moreover, \textit{dcm4che3} validator uses an undocumented XML file structure to determine the IOD structure and the mandatory Data Elements validation. The XML structure contemplates the IOD as the root element and the nested Data Elements. Furthermore, to enforce Value content validation against the VR attribute, Data Element can have an associated list of adaptable values that are useful for some attributes, for instance, the Patient’s Sex (0010,0040). The XML file also supports conditional elements by using the clauses And, If and Or, which allow the definition of dependencies.

In terms of patents, an invention from 1997 ~\cite{patent:5671353} proposes an object-oriented structure that includes a plurality of semantic definition and validation objects, and a method that semantically validates the DICOM message by passing them through the structure and comparing the DICOM message to the provided definitions. In 2001, another patent ~\cite{patent:20040205563} proposed a method for providing DICOM SR constraints within an XML document. To do so, the XML document was created containing DICOM SR constraints using declarative language. Moreover, a work from 2008 ~\cite{patent:20080071825} proposes a technique that employs a XML validation document with a set of constraints specified for   DICOM objects and makes use of them in validation processes.

Previously described validators are representative of the state-of-the-art in this field. They provide very useful functionalities but also have major limitations. First, these validators cannot resolve static preconditions that are dependent on the exam’s protocol, rather than on the IOD itself. In other words, conditions that require input from the user to know how to validate the DICOM file. An example is the condition "C Required if contrast media was used in this image”, which is present in many IODs. Secondly, the complexity of defining an entire configuration file for each IOD. This problem is aggravated by the first limitation since it creates the need to specify many configurations for the same IOD.

\section{Proposed Architecture}
\label{S:4}
\subsection{Overview}
\label{S:S1}
The proposed platform provides a set of functionalities supported by state-of-the-art solutions but solves the precondition problem by creating an engine that uses the input from the user to deal with the preconditions and load the required modules. Since defining an entire configuration file for an IOD is extremely complex, a modular approach was followed where each module of the IOD file is defined in a nested separated file. 
This is done by using an enhanced configuration interface that defines both IODs and Modules. This architecture allows re-use of the modules when creating new IOD definitions. 
Aiming to support programmers and research teams from academy and industry, in a collaborative way, it was decided to incorporate the validation software into an online platform where users can create and share their modules and definitions. Firstly, this creates a dynamic and synergistic environment easily able to adapt to the creation of new modalities, and secondly, reduces the need to define new configurations and simplifies use of the application.

\subsection{Functional Modules}
\label{S:Functional Modules}

The community-driven DICOM Validator has two main functionalities: Validation of DICOM files and support for the creation of description files, for Modality IODs and Modules, by the community. The description file is XML-based and specifies the Modules required for a given Modality IOD or what Attributes and respective Values are necessary for each Module. 

The documentation regarding the structure of the definition file for both files is included in the platform, and can be accessed at \url{https://bioinformatics.ua.pt/dicomvalidator/#/docs}.

\begin{figure}[!ht]
  \centering
    \includegraphics[width=.45\textwidth]{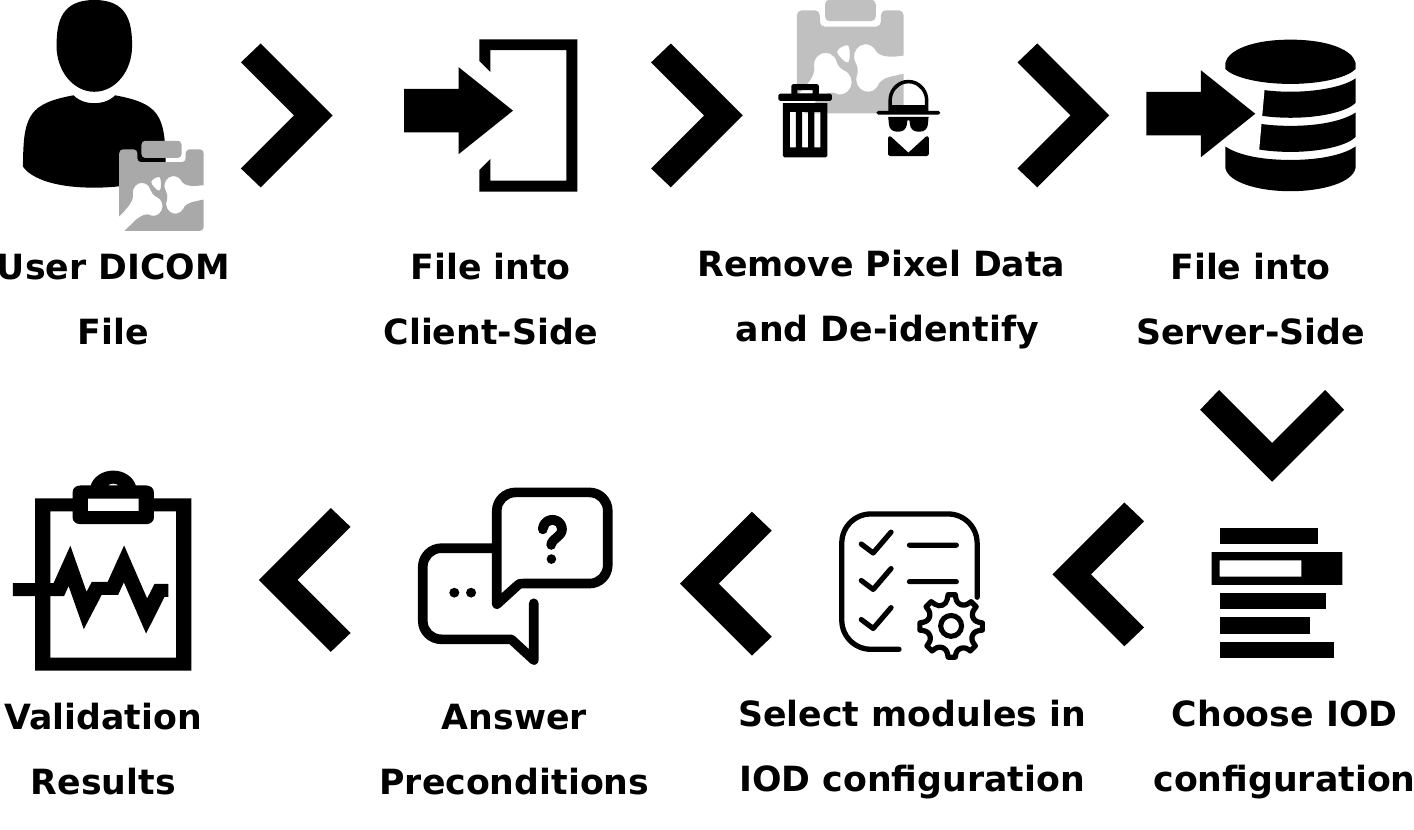}
 	\caption{\label{fig:Usecase1} Platform validation process.}
\end{figure}

Figure~\ref{fig:Usecase1} illustrates the validation use-case, where a user’s chosen file is uploaded to a remote server for validation. Before being sent, the file is locally modified (without overriding the user’s original one) to strip out pixel data and remove the patient's identifiable data for security and confidentiality reasons, as shown in Section \ref{S:Pixel Data Removal and De-Identification} 

\begin{figure}[!ht]
  \centering
    \includegraphics[width=0.4\textwidth]{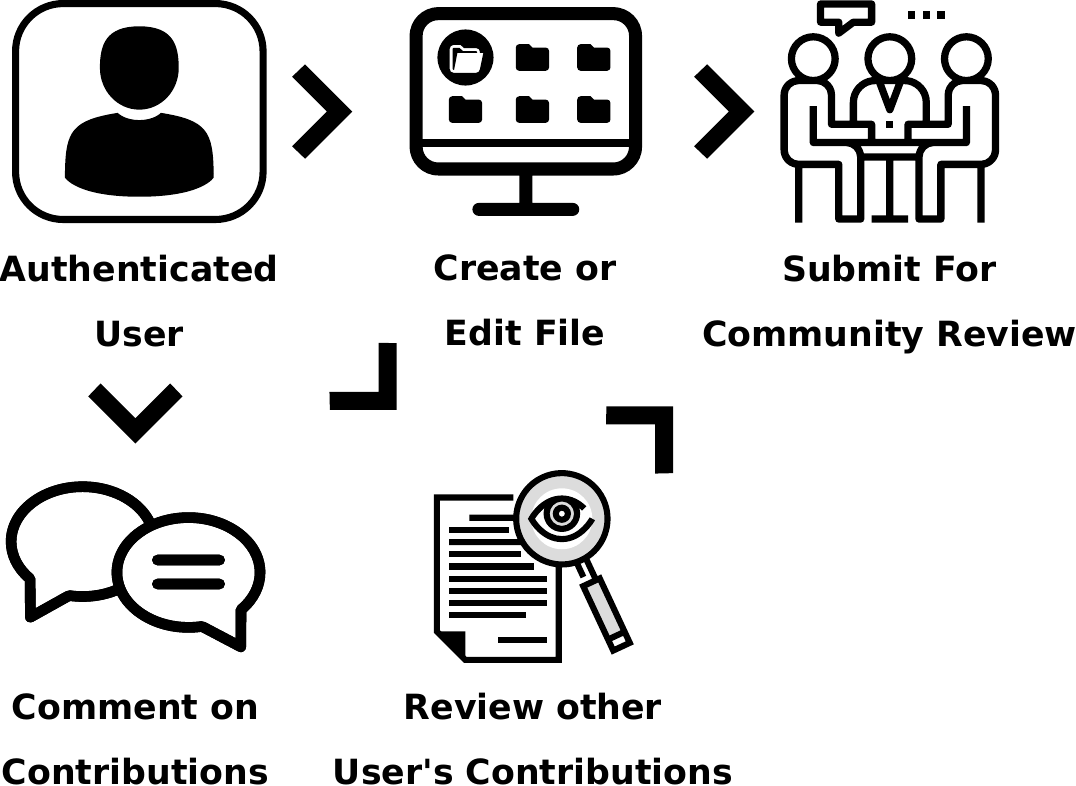}
 	\caption{\label{fig:Usecase2} Process of creation of description files by the community.}
\end{figure}

After receiving the file, the server loads the suitable description file for its Modality IOD and prompts the user to select which modules they want to validate. The user is then presented with several questions regarding the preconditions necessary for better validation results. A default answer to the questions can be provided.  Finally, the validation results are shown.
DICOM Validator provides a dashboard where users can follow up the validation processes. It also provides resources for the creation and editing of description files (Figure~\ref{fig:Usecase2}) using an embedded editor. These editing capabilities are required to make sure every description file is up-to-date and contains no errors. File editions are treated as contributions that can be commented on, and reviewed by the community before being accepted.

The use of dashboard features is only accessible to authenticated members to ensure the community concept and tracing of actions. 
Anyone can register and sign-in to the DICOM Validator platform. All authenticated users have access to the dashboard and can manually create or edit description files in the platform using a user-friendly editor, which are validated by the platform and can then be submitted as merge requests (pull requests). These merge requests are treated as contributions that can be commented on by the community before being accepted. Platform's administrator has the responsibility of accepting and merging these changes with the master repository to ensure the quality standards of the definition files.
The authentication system is handled by third-parties such as Google, GitHub, or Dropbox by using the \textit{OAuth} protocol\footnote{OAuth 2.0: https://oauth.net/2/}.

\subsection{Software layers}
\label{S:Software layers}

DICOM Validator consists of three software layers: a front-end, a back-end and a dashboard which encompasses a collaborative platform for editing description files.

The front-end is built using  \textit{Angular}\footnote{https://angular.io/} and \textit{Bootstrap} \footnote{http://getbootstrap.com/}, communicating with the back-end module through a REST API. It also includes a de-identification tool developed with \textit{dcmjs}\footnote{dcmjs: http://dcmjs.org/}.
The back-end module was built using \textit{Jetty} \footnote{http://www.eclipse.org/jetty/} to ensure communication, and additional Java Classes to perform validation, module loading, and user session logic. This module also stores the latest versions of the description files required by the validator.
Lastly, a modified version of \textit{Gitea}\footnote{https://gitea.io} was used to construct the dashboard. The changes include support for communication with the back-end and XML Editing using \textit{Xonomy}\footnote{http://www.lexiconista.com/xonomy/}.

\begin{figure}[!ht]
  \centering\includegraphics[width=0.40\textwidth]{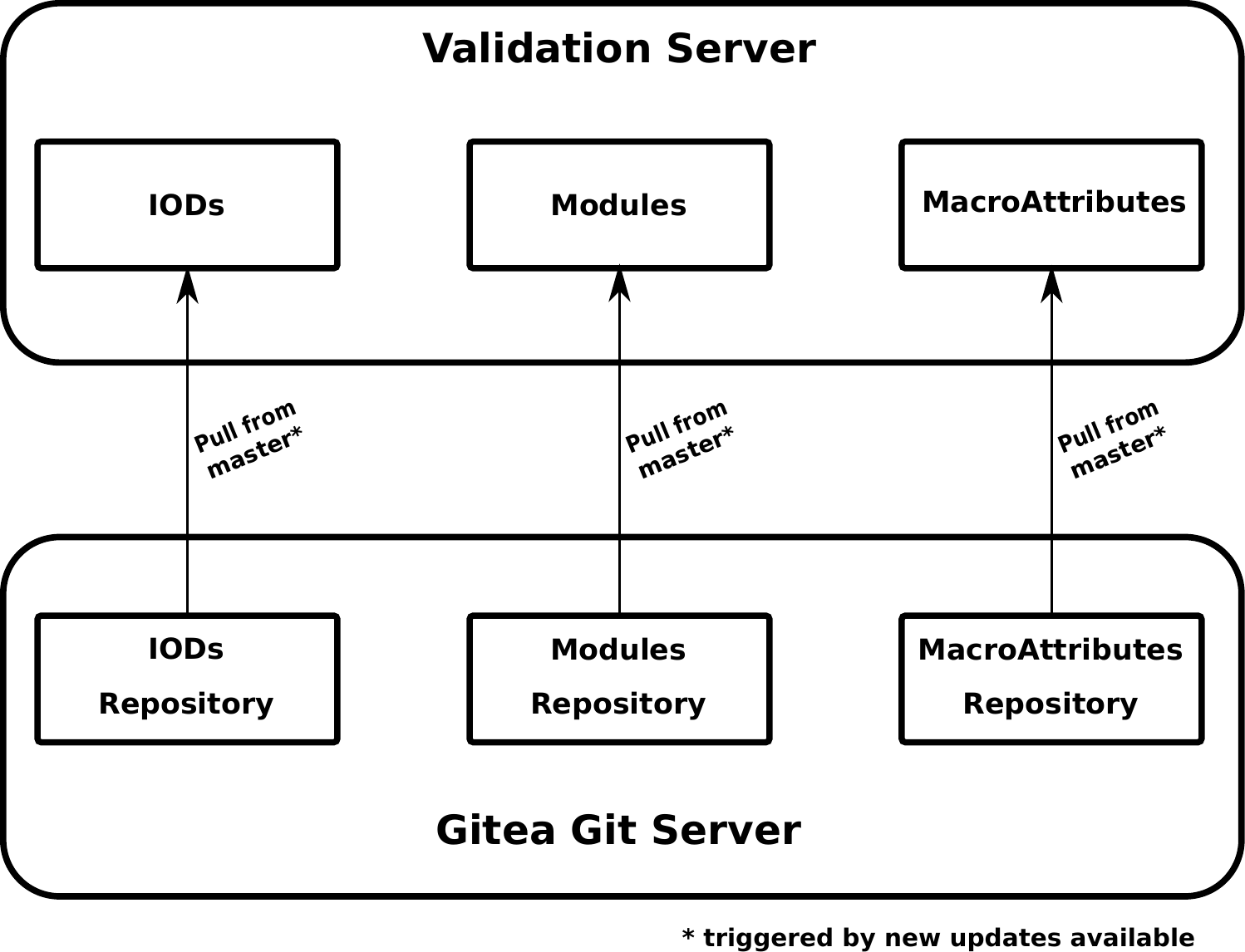}
 \caption{\label{fig:Comunitysync} Platform Creation or Editing of Modality IOD or Module Configurations Use-Case.}
\end{figure}

As Gitea is backed by a Git Server, that allows us to take advantage of this feature to keep the back-end description files up-to-date by syncing their respective folders with their repository’s master branch as shown in Figure~\ref{fig:Comunitysync}.

\subsection{Workflows}
\label{S:Workflows}
\subsubsection{Pixel Data Removal and De-Identification}
\label{S:Pixel Data Removal and De-Identification}

Typically, the DICOM file element requiring most storage space is the pixel data. For instance, the extreme case of DICOM objects for pathology imaging may require dozens of gigabytes~\cite{Singh2011StandardizationStandards.,2017AnRepositories}. Since the validation process does not request analysis of visual information, removal of pixel data on the client-side has the advantage of reducing the file’s upload time as well as the storage space required at the server. Furthermore, by removing pixel data, we are also eliminating any PHI that can be burned into images of some modalities such as in Ultrasound (US) and External-camera Photography~\cite{Silva2018ControlledArchives}.

DICOM standard PS3.15 Annex E provides the standard de-identification guidelines \cite{NEMA2018DICOME}. Table \ref{tab:StandardvsValidator} in the Supplementary Material provides comprehensive information regarding actions that the DICOM standard recommends for the Basic Profile attributes that we de-identified, and examples of what we provided as value for that attribute. However, our de-identification mechanism is not completely compliant with the standard, since we intended to create a fast mechanism that could easily replace the values of the attributes with dummy values consistent with the original attribute's value in terms of length and data type. This was done to ensure that we would not affect the DICOM object's validation regarding its structure. The unique objective is to use the data in the validation process, not the archiving or sharing of DICOM anonymized data.

The proposed platform uses \textit{dcmjs}, a Javascript cross-compilation of \textit{dcmtk}\footnote{dcmtk: http://dicom.offis.de/dcmtk.php.en}, to transform the input DICOM file into a representative XML file. The pixel data is removed in this transcoding process and the metadata is de-identified to ensure patient privacy and confidentiality~\cite{Douraki1999}. The de-identification process is done by reading the XML elements and replacing the fields containing the patient’s personal health information (PHI). The replacement (anonymous) sequence has the same length and domain type of original data, and is in accordance with the lists of adaptable values for that Data Element (see Section ~\ref{S:Validation Process}). For example, a \textit{PatientName} attribute with the content” \textit{DoeˆJohn}” would become” \textit{REMOVEDR}” and a \textit{PatientAge} with \textit{26} would become \textit{00}. Once this process is concluded, the XML file is transformed again into DICOM and passed to the server for validation.

\section{Validation Process}
\label{S:Validation Process}

Our web interface lies in a Model-View-Controller (MVC) paradigm and the framework works following the Service-Oriented Architecture (SOA), where the client-server communication is performed via Restless services. Our services are stateful and can be consumed by third party.
The services exposed by our API are the following:
\begin{itemize}
\item \textbf{\slash configure}: endpoint where files are uploaded to the platform and, as response, it is returned a validation ID and a set of modules and options.
\item \textbf{\slash validate}: to provide the answer to “/configure” service.  For instance, which modules will be validated, as well as the usage (or not) of the default answers for the preconditions. It returns the precondition questions or the file’s validation result if the user opted by the default answers to the preconditions.
\item \textbf{\slash result}: receives a validation ID and a response to the precondition questions (optional); and returns the file’s validation results.
\end{itemize}

The validation process is released on the server side. When a file reaches the application’s server, the DICOM metadata is read to obtain the \textit{SOPClassUID}, which is used to identify the medical imaging modality and select the respective description file to be used in the validation process. The selection of an appropriate description file is essential to obtain reliable results, since distinct DICOM modalities have different requirements.
An IOD description file is made by including the modules and preconditions that the user must answer. The preconditions can be placed either at the level of the IOD or the module depending on where they will be used (i.e., for inclusion of Modules or Data Elements, respectively). 
Listing~\ref{lst:xml1} shows part of the IOD description file for the CR Modality IOD. Included Modules are defined in the file by their Information Entity (IE), module name and use (Mandatory (M), Conditional (C) or User Defined (U)). A precondition is defined with an id name, a default value and a question that will be asked to the user during the validation process.

\begin{center}

\lstset{language=XML,morekeywords={IOD ,ie , define, include, If, \<, \/, \>, id, ie, value, question, module, usage, idref}}
\begin{lstlisting}[basicstyle=\scriptsize,showstringspaces=false,caption={Platform IOD configuration example for CR modality},label={lst:xml1}]
<IOD> 
 <define id="contrasMediaWasUsed" value="true" 
  question="Was contrast media used in this Image">
    
 <include ie="Patient" module="Patient" usage="M" />

 <include ie="Study" module="GeneralStudy" usage="M" />
 <include ie="Study" module="PatientStudy" usage="U" />

 <include ie="Series" module="GeneralSeries" usage="M" />

 <include ie="Image" module="GeneralImage" usage="M" />
    
 <include ie="Image" module="ContrastBolus" usage="M" >
  <If idref="contrasMediaWasUsed" />
 </include>
    
 <include ie="Image" module="SOPCommon" usage="M" />   
</IOD>
\end{lstlisting}
\end{center}

Besides preconditions, Module description files may include Macro Attribute tables and contain the Data Elements in accordance with the standard (Listing \ref{lst:xml2}). Each Data Element is categorized by a keyword, a tag, a VR, a VM, a type and the number of items (only if VR is a Sequence (SQ)). Furthermore, each Data Element can have an associated list of acceptable values.

\begin{center}
\lstset{language=XML,morekeywords={Module,define, DataElement, keyword,tag, vr, type, vm, Item, idref, Value, include, If, \<, \/, \>, id, ie, value, question, module, usage, idref}}
\begin{lstlisting}[basicstyle=\scriptsize,showstringspaces=false,caption={Platform Module example of a Definitions XML file for the Patient Module},label={lst:xml2}]
<Module>
    <define id="PatientIsAnAnimal" 
        question="Is patient an animal?" 
        default="false" />
	
    <define id="ResponsiblePersonIsPresent" 
        question="Is the responsible person present?" 
        default="false" />
	
    <DataElement 
        keyword="PatientName" 
        tag="00100010" 
        vr="PN" type="2" vm="1" />

    <DataElement 
        keyword="PatientID" 
        tag="00100020" 
        vr="LO" type="2" vm="1" />

    <include table="10-18" />

    <DataElement 
        keyword="PatientBirthDate" 
        tag="00100030" 
        vr="DA" type="2" vm="1" />
	
    <DataElement 
        keyword="PatientSex" 
        tag="00100040" 
        vr="CS" type="2" vm="1" >
            <Value>M</Value>
            <Value>F</Value>
            <Value>O</Value>
    </DataElement>

    ...
</Module>
\end{lstlisting}
\end{center}

Platform users may be questioned for two reasons: to select the modules to be validated and to answer preconditions. Depending on the user’s actions, modules may be excluded or interpreted in diverse ways.
Description file parsing is made using a custom XML handler based on \textit{dcm4che3’s SAXHandler} \footnote{https://github.com/dcm4che/dcm4che/blob/master/dcm4che-core/src/main/java/org/dcm4che3/data/IOD.java}and data validation is made using \textit{dcm4che3}’s validator. Since \textit{dcm4che3} validator natively ignores the validation of the Data Elements of type 3, it was necessary to adapt its behaviour to add support for validation of elements of this type. Moreover, the proposed validator extends the \textit{dcm4che3} validator by adding additional conditions, preconditions and a more modular approach to inclusion of the module in the IOD description files.
The results of the validation process are shown in 5 types of categories regarding the validation of each module:
\begin{itemize}
\item \textbf{Valid}, when all of the module’s content is in accordance with the standard;
\item \textbf{Warning}, when Data Elements of type 3 are not compliant with the standard;
\item \textbf{Skipped}, when the user chooses not to select a specific IOD module for validation;
\item \textbf{Unsatisfied Condition}, when a requested condition for validating a module was not met;
\item \textbf{Has Errors}, when Data Elements of type 1 and 2 are not in accordance with the standard.
\end{itemize}
All the messages exchanged between the client and server modules (e.g. user's answer to questions and results of validation process) are based on JSON object.

\section{Results}
\label{s:Results}

\subsection{Management Dashboard}
\label{Management Dashboard}
\begin{figure}[!ht]
  \includegraphics[width=0.45\textwidth]{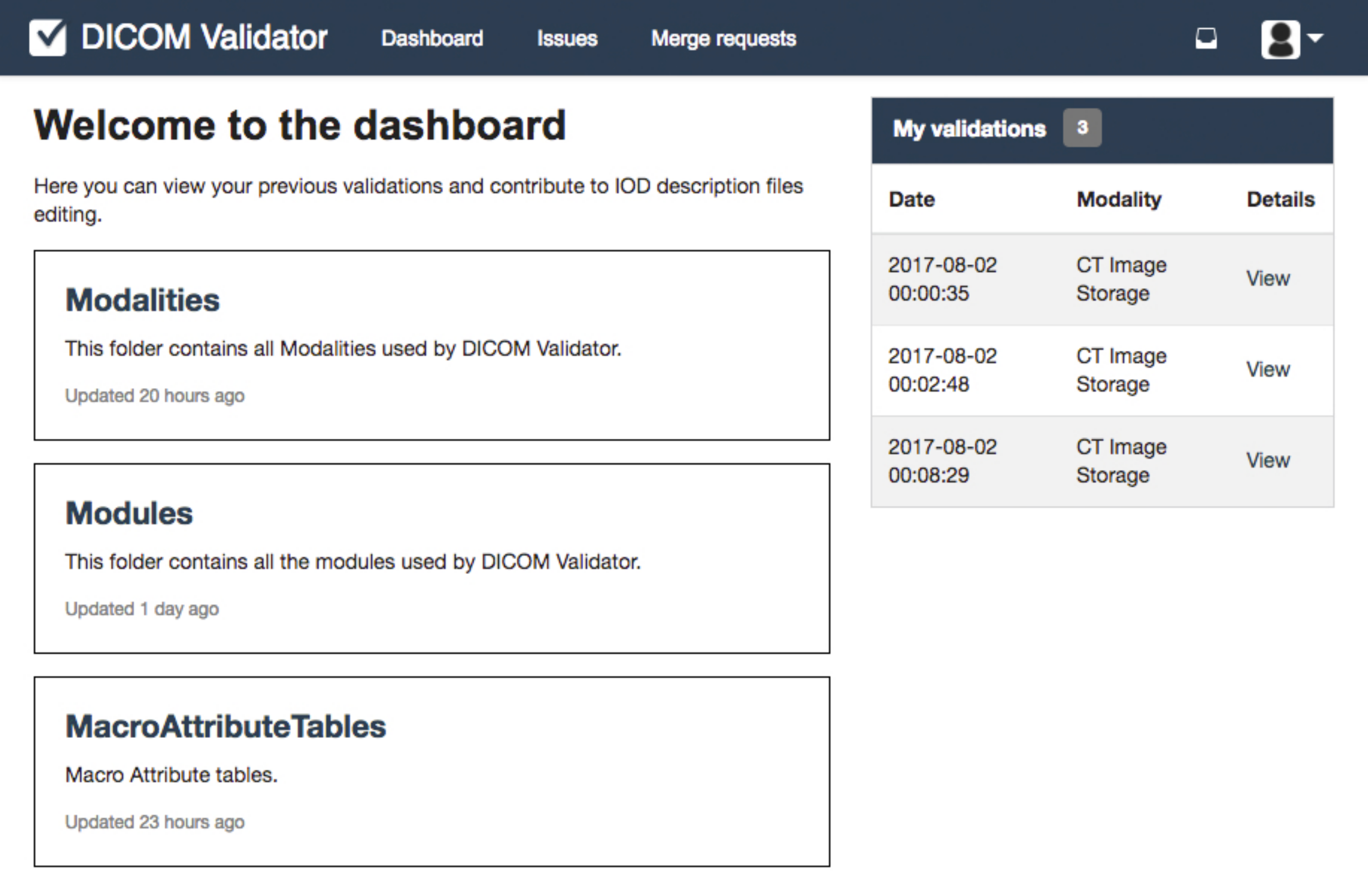}
 	\caption{\label{fig:dashboard}DICOM Validator Dashboard.}
\end{figure}

The Validator’s Dashboard page allows users to browse and edit the contents of Modules, Modality IODs and Macro Attribute Table repositories (Figure~\ref{fig:dashboard}). 

\begin{figure}[!ht]
  \centering\includegraphics[width=0.45\textwidth]{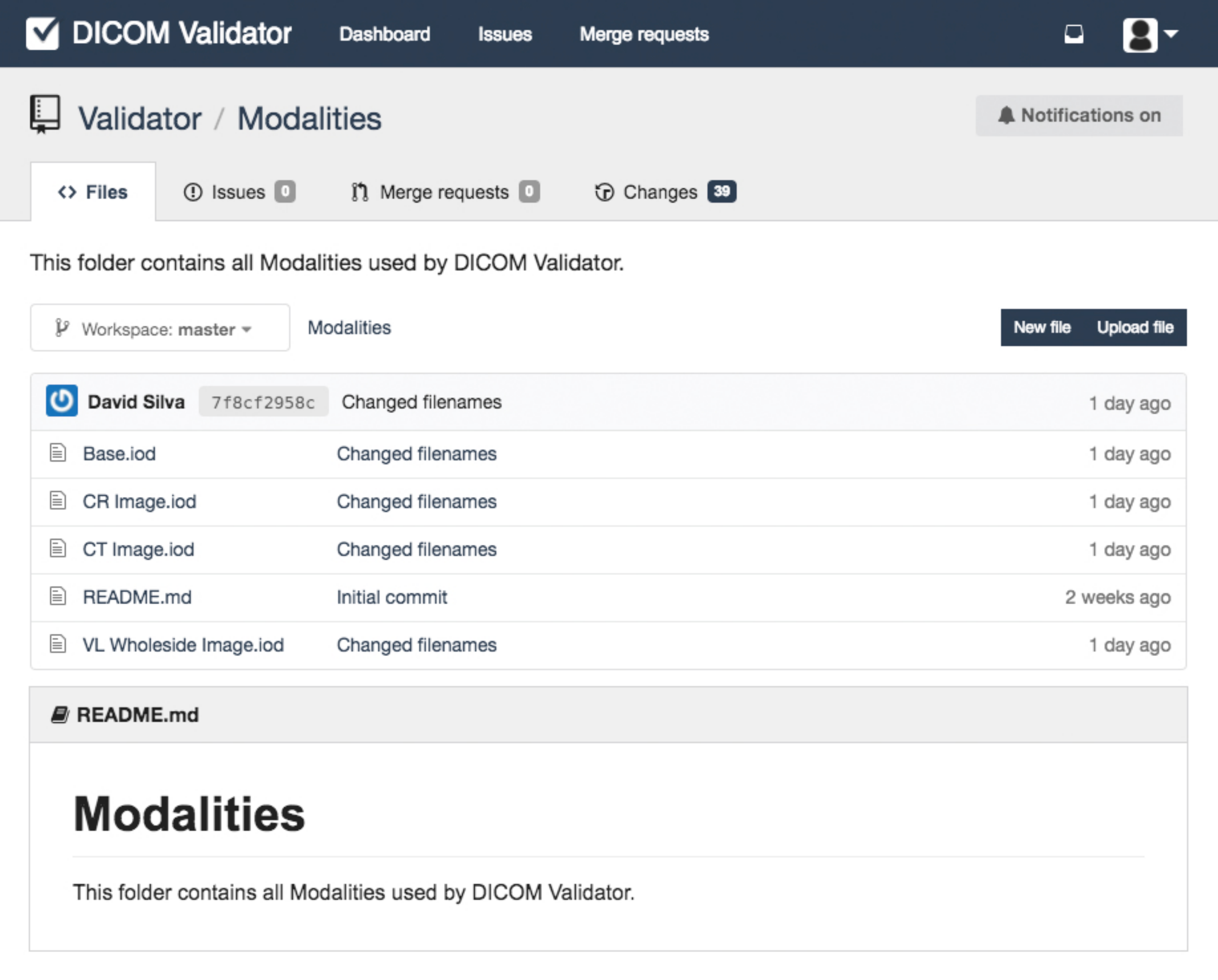}
 \caption{\label{fig:modalities} View of existing definition files for the Modality IODs. }
\end{figure}

The user can view the existing definition files for the Modality IODs (Figure~\ref{fig:modalities}) and Modules. The dashboard can show the user the history of their previous validations. Since the solution is back-end by a Git server, the platform can track the updating of description files by the different community members, accepting (merging) or rejecting the contribution requests, controlling in this way the quality of definition files used in production. Before submitting a new (file) contribution, the user must specify what changes were made by providing a title for the contribution and a description. The community can comment on the changes and suggest improvements, and the platform administrator can even approve the changes, merging the new code with the existing one in the master workspace and triggering the back-end validation service to update its description files to support the new ones.

\begin{figure}[!ht]
  \centering\includegraphics[width=0.45\textwidth]{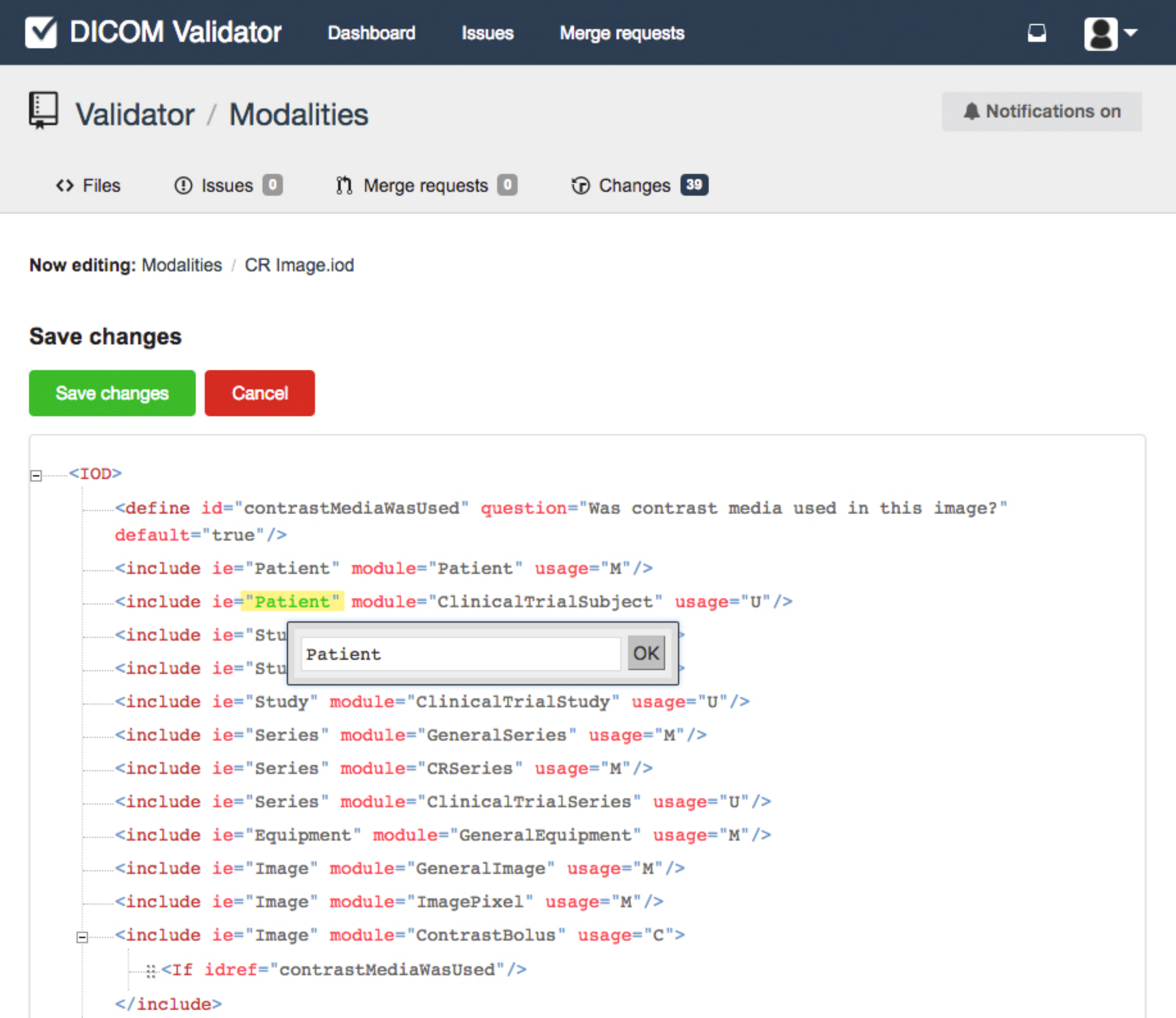}
 \caption{\label{fig:CRImageEdit} Edition of the definition file for the CR Modality IOD.}
\end{figure}

A description file can be edited in the platform using a user-friendly editor (Figure ~\ref{fig:CRImageEdit}).
This facilitates the creation of description files by automating several important tasks.
Besides editing, the user can “open issues” for situations that may be happening, and view changes made by other users.

After the editing is finished, the changed file is submitted to the back-end service, where the validator evaluates this new contribution and enumerates eventual errors found while parsing the description file. To do so, it first tries to identify the content as belonging to a file definition of a module or IOD. Once identified, it attempts to convert the XML to an IOD Module object or an IOD object, respectively. If the build fails, it is because the file structure is wrong, and an error is reported to the user. The user can then identify in which lines an error is present and correct them. If the validator ensures there are no errors, the description file can be submitted for community review.

\subsection{Validation Service}
\label{s:Validation Service}

Our proposed platform can be easily accessed via the URL \href{https://bioinformatics.ua.pt/dicomvalidator/}{\textbf{\textit{https://bioinformatics.ua.pt/dicomvalidator/}}}.

To evaluate the validation service, 3 DICOM files from distinct modalities were used and originated from a publicly available dataset:
\begin{itemize}
\item A Computed Radiography (CR)  modality, from Belarus Tuberculosis Portal (\textit{Belarus TB})  \footnote{Belarus TB: http://tuberculosis.by/}
\item A Computed Tomography (CT) from the \textit{DICOM Library} \footnote{DICOM Library: http://www.dicomlibrary.com/}
\item  A Whole Slide Imaging (WSI) from\textit{ DICOM WG26 -Pathology}\footnote{Nema: ftp://medical.nema.org/medical/dicom/DataSets/WG26/Hamamatsu/}
\end{itemize}

\begin{figure}
  \centering\includegraphics[height=.94\textheight]{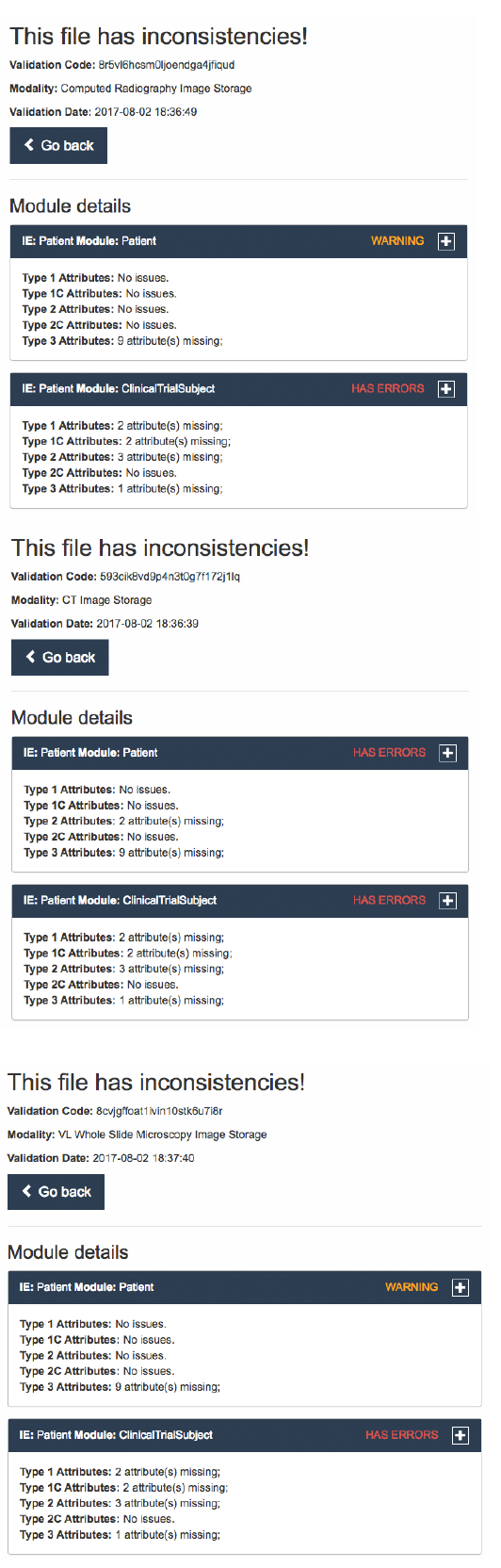}
 \caption{\label{fig:validation_results} Sample of the validation results obtained for the CR, CT and WSI files, showing for each file the errors occurred in each module ordered by the Data Element Type.}
\end{figure}

For each of the previous samples, the corresponding IOD definition and module files were created and made available in the DICOM Validator platform. Next, the files went through the pipeline described in Figure~\ref{fig:Usecase1}, where preconditions were answered as default. The validation results are presented in the Figure~\ref{fig:validation_results} and can be consulted by clicking on the validation codes shown in Table~\ref{tab:validation}.

\begin{table}[!th]
\centering
\caption{Validation ID of for each modality file validated.}
\label{tab:validation}
\begin{tabular}{c|c}
\textbf{Modality}    & \textbf{Validation Code} \\ \hline
CR & \href{https://bioinformatics.ua.pt/dicomvalidator/#/results/593cik8vd9p4n3t0g7f172j1lq}{8r5vl6hcsm0ljoendga4jfiqud}                \\ \hline
CT & \href{https://bioinformatics.ua.pt/dicomvalidator/#/results/8r5vl6hcsm0ljoendga4jfiqud}{593cik8vd9p4n3t0g7f172j1lq}                \\ \hline
WSI            & \href{https://bioinformatics.ua.pt/dicomvalidator/#/results/8cvjgffoat1ivin10stk6u7i8r}{8cvjgffoat1ivin10stk6u7i8r}  
\end{tabular}
\end{table}

Each validated module is classified using one of the categories presented in Section~\ref{S:Validation Process} (Valid, Has Errors, Warning, Skipped and Unsatisfied Condition). The results can be displayed as a summary or in a detailed view. The latter shows the attributes that are missing or have invalid values.
While analyzing other validators we noticed a lack of warnings regarding Type 3 attributes. Although they are legal, our line of reasoning was that their presence or absence as well as if they are in accordance with the standard should be displayed to the user. However, the display of such results may make it hard to distinguish important from unimportant results. As such, the platform provides a checkbox list that allows the users to select what types of error they wish to view after the validation.
An important aspect is that, after the validation, the user can go directly to \textit{DICOMLookup}\footnote{DICOMLookup: http://dicomlookup.com/}.  This provides the ability to link to context-sensitive information about each error. However, the \textit{DICOMLookup} service is not normative, nor necessarily maintained nor current. Therefore, we added a link directly to the CHTML online DICOM standard itself.

\section{Discussion}
\label{s:Discussion}

\subsection{Relevance of the Work}
The acceptance of DICOM standard by medical industry and researchers has brought to the market many devices that are sold as being DICOM conformant. However, in practice, the interoperability between different DICOM equipment is not always accomplished easily and completely. For instance, a study evaluated the DICOM conformance of cardiac X-ray Angiography objects from ten vendors using two validation tools and five visualization applications ~\cite{MaartenH.Baljon1999QualityImplementations}. The results showed that only two datasets were completely conformant with the DICOM standard and three contained serious errors that inhibited their correct visualization. Another important example was performed at the German Congress of Radiology 2006, where radiologists were invited to bring their CD for a short test against the CD specification published by German Society of Radiology, and 80\% of those “real world” CDs failed the test~\cite{ref1}.

\subsection{Comparison with State-of-the-Art and Solution Limitations}
\begin{table*}
\centering
\caption{Qualitative comparison analysis between related works.}
\label{tab:relatedworks}
\begin{tabular}{c|c|c|c|c|c|c}
Validation Software & \begin{tabular}[c]{@{}c@{}}Open \\ Source\end{tabular} & \begin{tabular}[c]{@{}c@{}}Web\\ Based\end{tabular} & De-identification & Preconditions & \begin{tabular}[c]{@{}c@{}}Consistency between \\ DICOM objects\end{tabular} & \begin{tabular}[c]{@{}c@{}}Update \\ Capability\end{tabular} \\ \hline
dicom3tools/dciodvfy &$\checkmark$ & $\times$ & $\times$ & $\times$ & $\times$ & $\times$ \\
dicom3tools/ dcentvfy &$\checkmark$ & $\times$ & $\times$ & $\times$ &$\checkmark$ & $\times$ \\
DVTK &$\checkmark$ & $\times$ & $\times$ & $\times$ & $\times$ & $\times$ \\
dcm4che3 &$\checkmark$ & $\times$ & $\times$ & $\times$ & $\times$ & $\times$ \\
DCMCHECK & $\times$ & $\times$ & $\times$ & $\times$ & $\times$ & $\times$ \\
DICOM Validator &$\checkmark$ &$\checkmark$ &$\checkmark$ &$\checkmark$ & $\times$ &$\checkmark$
\end{tabular}
\end{table*}

Table~\ref{tab:relatedworks} compares the different State-of-the-art  solutions with proposed DICOM Validator.

Currently, neither the DICOM Standards Committee nor MITA (NEMA) have any official tool or certification mechanism. However, evidence points to the fact that the existence of validation tools is fundamental to ensure interoperability between systems, since the dciodvfy tool and DVTk are used in IHE Connectathons for helping in the evaluation of systems interoperability~\cite{Clunie2015DICOMDciodvfy}.

As it can be observed in Table~\ref{tab:relatedworks}, most state-of-the-art solutions possess limitations that are covered by our solution, namely the fact that these validators are not web based, cannot resolve static preconditions and have a limited capability of keeping updated with the standard.
On the one hand, the fact that our system can run on a Web browser allows the user to perform simple tests without the need to deploy the system locally. On the other hand, the ability to resolve preconditions, as well as to host a community, provides users with more reliable validation results whilst maintaining the platform up to date with the standard.
The dciodvfy tool was Internet-deployed, but it is no longer available for the community. It used a CGI script on a web server that returned the output of an executable as 8Q8WKY-JLR4ZQif it runs in a command line. The tool did not support the de-identification of the images prior to uploading nor provided any mechanism for specifying predicates or to address real world conditional requirements.
Another important aspect is that the majority of these validators do not perform de-identification of PHI tags. Relatively to the platform trust model, there are no benefits from sending the identifiable object to the server. Independently of the trust in the organization that operates the server, it is always preferred to not share sensitive data with third entities, inclusively by legal restrictions~\cite{Sariyar2015SharingRequirements.}. The user feels more comfortable to use a service with this modus operandi. Moreover, the de-identification at client-side is an automatic process, not being a burden to the user. Finally, by removing the pixel data, the volume of data and consequent upload time is significantly reduced, which is fundamental to improve user experience.

Furthermore, there are some validators such as DVTK and dicom3tools/dcentvfy that support the evaluation of consistency between DICOM objects. Although this feature is important, it is worth noting that our validator was designed to validate individual DICOM objects and not to evaluate or compare the consistency between DICOM objects. As such, to perform such tasks one requires the use of the priorly mentioned validators.
  
 \subsection{Future Work}
Although we covered a large amount of requirements in order to create a solid, updated system, there are some aspects that we hope to tackle in future developments of our work.
 
Firstly, although the online service provided by the validator platform can be useful for some usage scenarios, developers may also request bulk verification that runs locally as well as its incorporation in automated regression testing frameworks. As such, one good improvement to this work will be the development of a command line extension that will consume the definition files created by the Validator's online community.

Secondly, a side-effect associated to the de-identification of identifiers’ attributes is that we are not able to validate them at server side. A possible solution is the user select to bypass the de-identification mechanism.
Removing pixel data is useful in terms of efficiency, since we are reducing the file’s upload time as well as the storage space required at the server. However, by doing so, proposed solution is failing to validate the actual pixel data, regarding its length and encoding. Currently, the platform maintains the number of frames of the file but does not consider the encoding and consequentially the length. As such, one possible improvement would be to upload pixel data at the expense of performance or to pass this information to the validator by extracting the information from the pixel data (7FE0,0010) prior to remove it at client side of the application, and then compare these values to the ones present in the metadata of the file.

Another important remark regarding the validator’s design was the use of preconditions. Although useful, answering all the precondition questions can be a quite tiresome and time-consuming task. On the other hand, leaving the selection to default can lead to the creation of many errors. As such, in future work, one could improve this work by creating a Management Dashboard feature that could let the user specify the granularity of the preconditions the user intends to respond.

Finally, another future improvement would be to provide users with a tool that could parse the DocBook XML of DICOM standard automatically and build our XML files. This tool would decrease the amount of user effort when creating new description files.

\section{Conclusion}
\label{s:Conclusion}

Digital medical imaging laboratories rely greatly on DICOM standard to ensure interoperability between different equipment. However, manufacturers' implementation of this standard may have non-conformities at several levels and medical staff may be responsible for data inconsistencies when entering data. The capacity to validate the quality of data and its compliance with the DICOM standard is a fundamental issue to avoid disruption in services.
This article presents an innovative community-driven web validation service for DICOM files. It runs in a common Web browser and can be safely used to validate real-world files since they are de-identified on the client side. The community can contribute to improving the platform by creating or editing description files used in the validation process. This means the platform can be updated whenever a new change is made to the DICOM standard or to resolve any issue that may produce unsatisfactory results and lead to unintended mistakes by the user. These contributions are always guided by a graphic interface tool and final verification is ensured before being submitted for the community's approval, meaning description files are less prone to errors. The results of file validation are saved on the database and can be accessed and shared by the user.
In conclusion, we propose a solution that solves problems faced by other state-of-the art validators and is also prepared to evolve along with the DICOM standard.

\section*{Acknowledgments}
Jorge Miguel Silva is funded by the ERDF – European Regional Development Fund through the Operational Programme for Competitiveness and Internationalisation – COMPETE 2020 Programme, and by National Funds through the FCT – Portugal (Funda\c{c}\~ao para a Ci\^encia e a Tecnologia) within project CMUP-ERI/ICT/0028/2014 SCREEN-DR. Tiago Marques Godinho is funded by FCT under grant agreement SFRH/BD/104647/2014.

\section*{References}
\bibliographystyle{IEEEtran}
\bibliography{Mendeley_Validation_Paper.bib}

\begin{thebibliography}{10}
\providecommand{\url}[1]{#1}
\csname url@samestyle\endcsname
\providecommand{\newblock}{\relax}
\providecommand{\bibinfo}[2]{#2}
\providecommand{\BIBentrySTDinterwordspacing}{\spaceskip=0pt\relax}
\providecommand{\BIBentryALTinterwordstretchfactor}{4}
\providecommand{\BIBentryALTinterwordspacing}{\spaceskip=\fontdimen2\font plus
\BIBentryALTinterwordstretchfactor\fontdimen3\font minus
  \fontdimen4\font\relax}
\providecommand{\BIBforeignlanguage}[2]{{%
\expandafter\ifx\csname l@#1\endcsname\relax
\typeout{** WARNING: IEEEtran.bst: No hyphenation pattern has been}%
\typeout{** loaded for the language `#1'. Using the pattern for}%
\typeout{** the default language instead.}%
\else
\language=\csname l@#1\endcsname
\fi
#2}}
\providecommand{\BIBdecl}{\relax}
\BIBdecl

\bibitem{Huang_2010}
\BIBentryALTinterwordspacing
H.~K. Huang, \emph{{PACS and Imaging Informatics: Basic Principles and
  Applications}}, 2nd~ed.\hskip 1em plus 0.5em minus 0.4em\relax Wiley, 2010.
  [Online]. Available: \url{http://books.google.pt/books?id=Pjjkyae_55oC}
\BIBentrySTDinterwordspacing

\bibitem{Creighton1999ASystems}
\BIBentryALTinterwordspacing
C.~Creighton, ``{A literature review on communication between picture archiving
  and communication systems and radiology information systems and/or hospital
  information systems},'' \emph{Journal of Digital Imaging}, vol.~12, no.~3,
  pp. 138--143, 8 1999. [Online]. Available:
  \url{http://link.springer.com/10.1007/BF03168632}
\BIBentrySTDinterwordspacing

\bibitem{St.Cyr2013AnStandards}
\BIBentryALTinterwordspacing
T.~J. St.~Cyr, ``{An overview of healthcare standards},'' in \emph{2013
  Proceedings of IEEE Southeastcon}.\hskip 1em plus 0.5em minus 0.4em\relax
  IEEE, 4 2013, pp. 1--5. [Online]. Available:
  \url{http://ieeexplore.ieee.org/document/6567436/}
\BIBentrySTDinterwordspacing

\bibitem{Mildenberger2002IntroductionStandard}
\BIBentryALTinterwordspacing
P.~Mildenberger, M.~Eichelberg, and E.~Martin, ``{Introduction to the DICOM
  standard},'' \emph{European Radiology}, vol.~12, no.~4, pp. 920--927, 4 2002.
  [Online]. Available: \url{http://link.springer.com/10.1007/s003300101100}
\BIBentrySTDinterwordspacing

\bibitem{Webb2003IntroductionImaging}
A.~A.~G. Webb, \emph{{Introduction to biomedical imaging}}.\hskip 1em plus
  0.5em minus 0.4em\relax Wiley-Interscience, 2003.

\bibitem{2003RealizationSystem}
\BIBentryALTinterwordspacing
``{Realization of integration and working procedure on digital hospital
  information system},'' \emph{Computer Standards {\&} Interfaces}, vol.~25,
  no.~5, pp. 529--537, 9 2003. [Online]. Available:
  \url{https://www.sciencedirect.com/science/article/pii/S0920548903000175}
\BIBentrySTDinterwordspacing

\bibitem{dreyer2006pacs}
\BIBentryALTinterwordspacing
K.~J. Dreyer, D.~S. Hirschorn, J.~H. Thrall, and A.~Mehta, \emph{{PACS: A Guide
  to the Digital Revolution}}.\hskip 1em plus 0.5em minus 0.4em\relax Springer
  New York, 2006. [Online]. Available:
  \url{https://books.google.pt/books?id=26PltGaHFFgC}
\BIBentrySTDinterwordspacing

\bibitem{dicomPart2Conformance}
{NEMA}, ``{Digital Imaging and Communications in Medicine (DICOM) Part 2 :
  Conformance},'' p. 320, 2018.

\bibitem{Horii2011DICOM}
\BIBentryALTinterwordspacing
S.~C. Horii, ``{DICOM},'' in \emph{Informatics in Medical Imaging}, S.~G.~L.
  George C.~Kagadis, Ed.\hskip 1em plus 0.5em minus 0.4em\relax CRC Press, 10
  2011, ch.~4, pp. 48--74. [Online]. Available:
  \url{https://www.taylorfrancis.com/books/e/9781439831366/chapters/10.1201%2Fb11382-6}
\BIBentrySTDinterwordspacing

\bibitem{Rosslyn2016}
{NEMA}, ``{Digital Imaging and Communications in Medicine (DICOM) Part 3 :
  Information Object Definitions},'' 2018.

\bibitem{Nema2018DigitalEncoding}
\BIBentryALTinterwordspacing
------, ``{Digital Imaging and Communications in Medicine (DICOM) Part 5: Data
  Structures and Encoding},'' p. 146, 2018. [Online]. Available:
  \url{http://dicom.nema.org/medical/dicom/current/output/pdf/part05.pdf}
\BIBentrySTDinterwordspacing

\bibitem{Oosterwijk2005DICOMBasics}
H.~Oosterwijk and P.~T. Gihring, \emph{{DICOM basics}}.\hskip 1em plus 0.5em
  minus 0.4em\relax OTech Inc, 2005.

\bibitem{dicom6}
{NEMA}, ``{Digital Imaging and Communications in Medicine ( DICOM ) Part 6 :
  Data Dictionary},'' 2018.

\bibitem{Hewett1997ConformanceObjects}
\BIBentryALTinterwordspacing
A.~J. Hewett, H.~Grevemeyer, A.~Barth, M.~Eichelberg, and P.~F. Jensch,
  ``{Conformance testing of DICOM image objects},'' in \emph{Proceedings of the
  SPIE, Volume 3035, p. 480-487 (1997).}, S.~C. Horii and G.~J. Blaine, Eds.,
  vol. 3035, 5 1997, pp. 480--487. [Online]. Available:
  \url{http://proceedings.spiedigitallibrary.org/proceeding.aspx?articleid=921948}
\BIBentrySTDinterwordspacing

\bibitem{NEMA2018DigitalInterchange}
\BIBentryALTinterwordspacing
{NEMA}, ``{Digital Imaging and Communications in Medicine (DICOM) Part 10:
  Media Storage and File Format for Media Interchange​},'' 2018. [Online].
  Available:
  \url{http://dicom.nema.org/medical/dicom/current/output/pdf/part10.pdf}
\BIBentrySTDinterwordspacing

\bibitem{Potter2007}
\BIBentryALTinterwordspacing
G.~Potter, R.~Busbridge, M.~Toland, and P.~Nagy, ``{Mastering DICOM with
  DVTk},'' \emph{Journal of Digital Imaging}, vol.~20, no.~1, pp. 47--62, 11
  2007. [Online]. Available: \url{https://doi.org/10.1007/s10278-007-9057-0}
\BIBentrySTDinterwordspacing

\bibitem{patent:5671353}
\BIBentryALTinterwordspacing
T.~M. B. C. C. C. M. D. G.~T. Tian Helen He~(Plano, ``{Method for validating a
  digital imaging communication standard message},'' p.~27, 1997. [Online].
  Available: \url{http://www.freepatentsonline.com/5671353.html}
\BIBentrySTDinterwordspacing

\bibitem{patent:20040205563}
\BIBentryALTinterwordspacing
N.~Y.~U. Lee Kwok Pun~(Flushing, ``{Specifying DICOM semantic constraints in
  XML},'' 2001. [Online]. Available:
  \url{http://www.freepatentsonline.com/y2004/0205563.html}
\BIBentrySTDinterwordspacing

\bibitem{patent:20080071825}
\BIBentryALTinterwordspacing
N.~H.~U. Guo Dongbai~(Nashua, ``{Techniques for checking whether a complex
  digital object conforms to a standard},'' p.~35, 2008. [Online]. Available:
  \url{http://www.freepatentsonline.com/y2008/0071825.html}
\BIBentrySTDinterwordspacing

\bibitem{Singh2011StandardizationStandards.}
\BIBentryALTinterwordspacing
R.~Singh, L.~Chubb, L.~Pantanowitz, and A.~Parwani, ``{Standardization in
  digital pathology: Supplement 145 of the DICOM standards.}'' \emph{Journal of
  pathology informatics}, vol.~2, p.~23, 2011. [Online]. Available:
  \url{http://www.ncbi.nlm.nih.gov/pubmed/21633489
  http://www.pubmedcentral.nih.gov/articlerender.fcgi?artid=PMC3097525}
\BIBentrySTDinterwordspacing

\bibitem{2017AnRepositories}
\BIBentryALTinterwordspacing
``{An efficient architecture to support digital pathology in standard medical
  imaging repositories},'' \emph{Journal of Biomedical Informatics}, vol.~71,
  pp. 190--197, 7 2017. [Online]. Available:
  \url{http://www.sciencedirect.com/science/article/pii/S1532046417301326}
\BIBentrySTDinterwordspacing

\bibitem{Silva2018ControlledArchives}
\BIBentryALTinterwordspacing
J.~M. Silva, E.~Pinho, E.~Monteiro, J.~F. Silva, and C.~Costa, ``{Controlled
  searching in reversibly de-identified medical imaging archives},''
  \emph{Journal of Biomedical Informatics}, vol.~77, pp. 81--90, 1 2018.
  [Online]. Available:
  \url{https://www.sciencedirect.com/science/article/pii/S1532046417302721?via%3Dihub}
\BIBentrySTDinterwordspacing

\bibitem{NEMA2018DICOME}
\BIBentryALTinterwordspacing
{NEMA}, ``{DICOM standard PS3.15 Annex E},'' 2018. [Online]. Available:
  \url{http://dicom.nema.org/medical/dicom/current/output/html/part15.html#chapter_E}
\BIBentrySTDinterwordspacing

\bibitem{Douraki1999}
\BIBentryALTinterwordspacing
T.~Douraki, ``{Ethical and Legal Dimensions of Medical Confidentiality in
  European Law of Human Rights},'' in \emph{Manage or Perish?}\hskip 1em plus
  0.5em minus 0.4em\relax Boston, MA: Springer US, 1999, pp. 421--427.
  [Online]. Available:
  \url{http://link.springer.com/10.1007/978-1-4615-4147-9_50}
\BIBentrySTDinterwordspacing

\bibitem{MaartenH.Baljon1999QualityImplementations}
\BIBentryALTinterwordspacing
{Maarten H. Baljon}, {Maarten G. Gerritsen}, {Marco Eichelberg}, and {Peter
  Jensch}, ``{Quality Control using Automated Validation Tools can Improve
  Interoperability of DICOM Implementations},'' Barcelona, p.~3, 1999.
  [Online]. Available:
  \url{http://www.uni-kiel.de/Kardiologie/dicom/1999/dicom_validation.html#p1}
\BIBentrySTDinterwordspacing

\bibitem{ref1}
\BIBentryALTinterwordspacing
M.~Onkena, M.~Eichelberga, J.~Riesmeierb, and P.~Mildenbergerc, ``{Image
  distribution and integration strategies - Exchange of radiological images on
  DICOM CD: a survey of the state of technology in Germany},''
  \emph{International Journal of Computer Assisted Radiology and Surgery},
  vol.~2, no.~1, pp. 290--292, 6 2007. [Online]. Available:
  \url{https://doi.org/10.1007/s11548-007-0101-9}
\BIBentrySTDinterwordspacing

\bibitem{Clunie2015DICOMDciodvfy}
\BIBentryALTinterwordspacing
D.~Clunie, ``{DICOM Validator - dciodvfy},'' 2015. [Online]. Available:
  \url{http://www.dclunie.com/dicom3tools/dciodvfy.html}
\BIBentrySTDinterwordspacing

\bibitem{Sariyar2015SharingRequirements.}
M.~Sariyar, I.~Schluender, C.~Smee, and S.~Suhr,
  ``\BIBforeignlanguage{eng}{{Sharing and Reuse of Sensitive Data and Samples:
  Supporting Researchers in Identifying Ethical and Legal Requirements.}}''
  \emph{\BIBforeignlanguage{eng}{Biopreservation and biobanking}}, vol.~13,
  no.~4, pp. 263--270, 8 2015.

\end{thebibliography}
\onecolumn
\renewcommand*{\arraystretch}{1.4}
\section*{Supplementary Material}
\pagenumbering{gobble}
\begin{center}
\begin{longtable}{|m{3cm}|l|m{1.5cm}|m{1.5cm}|m{9cm}|}
\caption[LoF entry]{List of DICOM attributes de-identified by the Validator. Actions recommended by the Standard for the Basic Profile in Comparison with our approach.\label{tab:StandardvsValidator}
\textbf{Notation:}\\ \textbf{U} -- Replace with a non-zero length UID that is internally consistent within a set of Instances\\ 
\textbf{X} -- Remove\\
\textbf{Z} -- Replace with a zero-length value, or a non-zero length value that may be a dummy value and consistent with the VR\\
\textbf{X/D} -- X unless D is required to maintain IOD conformance (Type 3 versus Type 1)\\
\textbf{X/Z/D} -- X unless Z or D is required to maintain IOD conformance (Type 3 versus Type 2 versus Type 1)
}\\
\hline
\textbf{Attribute Name}                 & \textbf{Tag} & \textbf{\begin{tabular}[c]{@{}l@{}}Action in\\DICOM\\ Standard\end{tabular}} & \textbf{\begin{tabular}[c]{@{}l@{}}Action in\\  Validator\end{tabular}} & \textbf{\begin{tabular}[c]{@{}l@{}}DICOM Validator's Example\\ of Replacement\end{tabular}} \\ \hline
Instance creator UID                    & (0008,0014)  & U                                                                           & Z                                                                       & 0.0.00.0.0000.0.0.0.00000000000000.00000000.000000.000                                   \\ \hline
SOP instance UID                        & (0008,0018)  & U                                                                           & Z                                                                       & 0.0.00.0.0000.0.0.0.00000000000000.00000000.000000.000                                      \\ \hline
Accession number                        & (0008,0050)  & Z                                                                           & Z                                                                       & REMOVED                                                                                     \\ \hline
Institution name                        & (0008,0080)  & X/Z/D                                                                       & Z                                                                       & REMOVED                                                                                     \\ \hline
Institution address                     & (0008,0081)  & X                                                                           & Z                                                                       & REMOVED                                                                                     \\ \hline
Referring physician's name              & (0008,0090)  & Z                                                                           & Z                                                                       & REMOVED                                                                                     \\ \hline
Referring physician's address           & (0008,0092)  & X                                                                           & Z                                                                       & REMOVED                                                                                     \\ \hline
Referring physician's telephone numbers & (0008,0094)  & X                                                                           & Z                                                                       & REMOVED                                                                                     \\ \hline
Station name                            & (0008,1010)  & X/Z/D                                                                       & Z                                                                       & REMOVED                                                                                     \\ \hline
Study description                       & (0008,1030)  & X                                                                           & Z                                                                       & REMOVED                                                                                     \\ \hline
Series description                      & (0008,103E)  & X                                                                           & Z                                                                       & REMOVED                                                                                     \\ \hline
Institutional\newline department name           & (0008,1040)  & X                                                                           & Z                                                                       & REMOVED                                                                                     \\ \hline
Physician(s) of\newline record                  & (0008,1048)  & X                                                                           & Z                                                                       & REMOVED                                                                                     \\ \hline
Performing\newline physicians' name             & (0008,1050)  & X                                                                           & Z                                                                       & REMOVED                                                                                     \\ \hline
Name of physician(s) reading study      & (0008,1060)  & X                                                                           & Z                                                                       & REMOVED                                                                                     \\ \hline
Operators' name                         & (0008,1070)  & X/Z/D                                                                       & Z                                                                       & REMOVED                                                                                     \\ \hline
Admitting diagnoses description         & (0008,1080)  & X                                                                           & Z                                                                       & REMOVED                                                                                     \\ \hline
Referenced SOP\newline instance UID             & (0008,1155)  & U                                                                           & Z                                                                       & 0.0.00.0.0000.0.0.0.00000000000000.00000000.000000.000                                      \\ \hline
Derivation\newline description                  & (0008,2111)  & X                                                                           & Z                                                                       & REMOVED                                                                                     \\ \hline
Patient's name                          & (0010,0010)  & Z                                                                           & Z                                                                       & REMOVED                                                                                     \\ \hline
Patient ID                              & (0010,0020)  & Z                                                                           & Z                                                                       & REMOVED                                                                                     \\ \hline
Patient's birth date                    & (0010,0030)  & Z                                                                           & Z                                                                       & 20070304                                                                                    \\ \hline
Patient's birth time                    & (0010,0032)  & X                                                                           & Z                                                                       & 20000101000000.000000\&0000                                                                 \\ \hline
Patient's sex                           & (0010,0040)  & Z                                                                           & Z                                                                       & O                                                                                           \\ \hline
Other patient Ids                       & (0010,1000)  & X                                                                           & Z                                                                       & 0                                                                                           \\ \hline
Other patient names                     & (0010,1001)  & X                                                                           & Z                                                                       & REMOVED                                                                                     \\ \hline
Patient's age                           & (0010,1010)  & X                                                                           & Z                                                                       & 00                                                                                          \\ \hline
Patient's size                          & (0010,1020)  & X                                                                           & Z                                                                       & 0                                                                                           \\ \hline
Patient's weight                        & (0010,1030)  & X                                                                           & Z                                                                       & 0                                                                                           \\ \hline
Medical\newline record locator                  & (0010,1090)  & X                                                                           & Z                                                                       & REMOVED                                                                                     \\ \hline
Ethnic group                            & (0010,2160)  & X                                                                           & Z                                                                       & REMOVED                                                                                     \\ \hline
Occupation                              & (0010,2180)  & X                                                                           & Z                                                                       & REMOVED                                                                                     \\ \hline
Additional patient's history            & (0010,21B0)  & X                                                                           & Z                                                                       & REMOVED                                                                                     \\ \hline
Patient comments                        & (0010,4000)  & X                                                                           & Z                                                                       & REMOVED                                                                                     \\ \hline
Device serial number                    & (0018,1000)  & X/Z/D                                                                       & Z                                                                       & 0                                                                                           \\ \hline
Protocol name                           & (0018,1030)  & X/D                                                                         & Z                                                                       & REMOVED                                                                                     \\ \hline
Study instance UID                      & (0020,000D)  & U                                                                           & Z                                                                       & 0.0.00.0.0000.0.0.0.00000000000000.00000000.000000.000                                      \\ \hline
Series instance UID                     & (0020,000E)  & U                                                                           & Z                                                                       & 0.0.00.0.0000.0.0.0.00000000000000.00000000.000000.000                                      \\ \hline
Study ID                                & (0020,0010)  & Z                                                                           & Z                                                                       & 0                                                                                           \\ \hline
Frame of reference UID                  & (0020,0052)  & U                                                                           & Z                                                                       & 0.0.00.0.0000.0.0.0.00000000000000.00000000.000000.000                                      \\ \hline
Synchronization frame of reference UID  & (0020,0200)  & U                                                                           & Z                                                                       & 0.0.00.0.0000.0.0.0.00000000000000.00000000.000000.000                                      \\ \hline
Image comments                          & (0020,4000)  & X                                                                           & Z                                                                       & REMOVED                                                                                     \\ \hline
UID                                     & (0040, A124) & U                                                                           & Z                                                                       & 0.0.00.0.0000.0.0.0.00000000000000.00000000.000000.000                                      \\ \hline
Storage media\newline file-set UID              & (0088,0140)  & U                                                                           & Z                                                                       & 0.0.00.0.0000.0.0.0.00000000000000.00000000.000000.000                                      \\ \hline
Referenced frame\newline of reference UID       & (3006,0024)  & U                                                                           & Z                                                                       & 0.0.00.0.0000.0.0.0.00000000000000.00000000.000000.000                                      \\ \hline
Related frame of\newline reference UID          & (3006,00C2)  & U                                                                           & Z                                                                       & 0.0.00.0.0000.0.0.0.00000000000000.00000000.000000.000                                      \\ \hline
\end{longtable}
\end{center}

\twocolumn

\end{document}